\pdfoutput=1

\documentclass[12pt, a4paper]{article}

\usepackage{amsmath}
\usepackage{amsfonts}
\usepackage{amssymb}
\usepackage{bbm}
\usepackage{verbatim}
\usepackage{dsfont}
\usepackage{booktabs}
\usepackage{slashed}
\usepackage{bm}

\usepackage{epsfig}
\usepackage{color}
\usepackage[table]{xcolor}
\usepackage{graphicx}
\usepackage[]{caption}
\usepackage[listofformat=empty,subrefformat=empty]{subfig}

\usepackage{float}
\usepackage{overpic}

\usepackage{enumerate}
\usepackage{hhline}
\usepackage{multirow}

\usepackage{cite}
\usepackage{a4wide}
\usepackage{bbold}

\usepackage{geometry}
\renewcommand{\baselinestretch}{1.2}

\newcommand{\A}{\mathcal{A}}
\newcommand{\T}{\mathcal{T}}

\newcommand{\pd}{\partial}

\begin{document}

\thispagestyle{empty}

\begin{flushright}
DESY 22-059
\end{flushright}
\vskip .8 cm
\begin{center}
  {\Large {\bf The cosmological constant as a boundary term}}\\[12pt]

\bigskip
\bigskip 
{
{\bf{Wilfried Buchm\"uller$^\dagger$}\footnote{E-mail:
    wilfried.buchmueller@desy.de}} and
{\bf{Norbert Dragon$^\ast$}\footnote{E-mail: dragon@itp.uni-hannover.de}}
}
\bigskip\\[0pt]
\vspace{0.23cm}
{\it $^\dagger$ Deutsches Elektronen-Synchrotron DESY, 22607 Hamburg, Germany \\
  \vspace{0.2cm}
  $^\ast$ Institut für Theoretische Physik, Leibniz Universität
  Hannover,\\ 30167 Hannover, Germany
}
\\[20pt] 
\bigskip
\end{center}

\date{\today}

\begin{abstract}
\noindent
We compare the path integral for transition functions in unimodular
gravity and in general relativity.  In unimodular gravity the
cosmological constant is a property of states that are specified at
the boundaries whereas in general relativity the cosmological constant
is a parameter of the action. Unimodular gravity with a nondynamical
background spacetime volume element has a time variable that is
canonically conjugate to the cosmological constant. Wave functions
depend on time and satisfy a Schr\"odinger equation. On the contrary,
in the covariant version of unimodular gravity with a 3-form gauge
field, proposed by Henneaux
and Teitelboim, wave functions are time independent and satisfy a
Wheeler-DeWitt equation, as in general relativity. The 3-form gauge
field integrated over spacelike hypersurfaces becomes a ``cosmic
time'' only in the semiclassical approximation. In unimodular gravity
the smallness of the observed cosmological constant has to be
explained as a property of the initial state.
\end{abstract}

\newpage 
\setcounter{page}{2}
\setcounter{footnote}{0}
{
  \renewcommand{\baselinestretch}{1}\tableofcontents}

\section{Introduction}
\label{sec:introduction}
The origin and interpretation as well as the observed value of the
cosmological constant presents a puzzle of particle physics and cosmology
\cite{Weinberg:1988cp}. In particular the seemingly huge contribution
of zero-point energies is often
considered to be a severe fine-tuning problem. It is therefore
suggestive that the cosmological constant is just an integration
constant, rather than a fundamental parameter, in a version of Einstein's theory 
where the volume element $\sqrt{g}$
is fixed. This has been noticed
from time to time
\cite{Anderson:1971pn,vanderBij:1981ym,Wilczek:1983as,Zee:1983jg,Buchmuller:1988yn,Buchmuller:1988wx}
and has led to a canonical theory of quantum gravity \cite{Henneaux:1989zc ,Unruh:1988in,Unruh:1989db}.

Unimodular gravity (UG) can be defined by imposing $\sqrt{g} = \omega$ as a
constraint, where $\omega$ is a nondynamic background volume
element. One often chooses $\sqrt{g}=1$, hence the name unimodular gravity.
The background volume element breaks the invariance of general
relativity (GR) under general diffeomorphisms to
the invariance under volume preserving diffeomorphisms.
Nevertheless, the classical theory is equivalent to Einstein gravity
except for the cosmological constant which now appears as an
integration constant. This feature also arises in a generally
covariant theory with a 3-form gauge field, which was obtained by Henneaux and
Teitelboim in an analysis of unimodular gravity as a constrained
Hamiltonian system \cite{Henneaux:1989zc}.
Note that 3-form gauge
fields can also contribute to the cosmological constant by vacuum
expectation values of their 4-form field strengths
\cite{Duff:1980qv,Aurilia:1980xj,Henneaux:1984ji}.
Introducing further gauge fields also Newton's constant can become an
integration constant \cite{Jirousek:2020vhy,Kaloper:2022oqv}.

In a theory with invariance only under volume preserving
diffeomorphisms the conformal factor of the metric,
$\sigma = \frac{1}{2}\ln{(\sqrt{g})}$, is an ordinary scalar field
that can have arbitrary kinetic term and potential.
However, its couplings may be restricted by additional symmetries such as scale
invariance. In this way it plays a prominent role in Higgs-dilaton theories;
see, for example,
\cite{Wetterich:1987fm,Buchmuller:1988cj,Shaposhnikov:2008xb,Copeland:2021qby}.

During the past years quantum effects in
UG have been studied in detail, and there has been a
still ongoing debate whether or not UG and GR are equivalent as
quantum theories. The investigations include
semiclassical calculations \cite{Fiol:2008vk},  the quantum effective
action \cite{Smolin:2009ti}, the renormalization group flow
\cite{Eichhorn:2013xr,Saltas:2014cta,deBrito:2020rwu}, the quantum
equivalence of UG and GR \cite{Padilla:2014yea}, quantum corrections to
the cosmological constant \cite{Alvarez:2015pla,Alvarez:2015sba},
the path integral in the Hamiltonian formalism
\cite{Bufalo:2015wda,deLeonArdon:2017qzg,Percacci:2017fsy} and the
computation of one-loop divergencies \cite{Herrero-Valea:2020xaq}.
Recently, significant progress has been made in the BRST
quantization of UG as well as GR in the unimodular gauge
\cite{Baulieu:2020obv,Kugo:2021bej,Kugo:2022iob,Kugo:2022dui}.
 It is perhaps not surprizing that at present
 there is no consensus on how to precisely define 
 unimodular quantum gravity, and it is far from clear what the
 differences to ordinary quantum gravity are.
 
 In the following we shall attempt to compare the quantum theories of
 GR and the two versions of UG. The
 comparison will be based on the path integral for transition
 amplitudes. The main difference is that in UG 
 the cosmological constant enters as a boundary term, i.e., as a
 property of states, whereas in GR it is a parameter of the action.
 GR and the Henneaux-Teitelboim
 version of UG are generally covariant. Hence, there
 is no notion of time on which wave functions could depend. On the other hand,
 in UG with a nondynamical background volume element canonical
 quantization is possible and wave functions do depend on time.

 The paper is organized as follows. After a general discussion of the
 path integral and the Henneaux-Teitelboim action in Section~2 we analyze
 the path integral for unimodular gravity in Sections~3 and 4, with emphasis on
 the boundary terms. Wave functions are briefly considered in
 Section~5. We conlude in Section~6. BRST
 quantization of general relativity in unimodular gauge is discussed
 in the appendix.

\section{The path integral in quantum gravity}
\label{sec:pathintegral}
A natural starting point for quantizing gravity is the path
integal (see, for example, \cite{Hawking:1979ig,Hartle:1983ai}). To obtain an expression for the amplitude one has to
identify dynamical variables and study their ``time evolution''.
As a first step one introduces a ``time function'' $t(x)$
that provides a foliation of a hyperbolic spacetime manifold
$\mathcal{M}$ into spacelike 3-surfaces $\Sigma_t$.
One can then define transition amplitudes between states corresponding
to different configurations of the gravitational field on
3-surfaces of different ``parametric time'' $t$.
For simplicity, we shall restrict our discussion to 
compact 3-surfaces.

Einstein's equations for the gravitational field are obtained from the 
action\footnote{The volume form is given by
  $\epsilon = \frac{1}{4!}\sqrt{g}\epsilon_{\alpha\beta\gamma\delta}
dx^\alpha dx^\beta dx^\gamma dx^\delta$, where $g = -
\text{det}{g_{\alpha\beta}}$, and $\epsilon_{\alpha\beta\gamma\delta}$
is the Levi-Civita tensor density with
$\epsilon_{0123} = 1$.
$\tilde{\epsilon}$ is the induced volume form on $\partial\mathcal{M}$.
We
work in units $16\pi G_N = 1$.}
\begin{equation}\label{hilbert}
  S[g] = \int_{\mathcal{M}} R \epsilon + 2\int_{\partial\mathcal{M}} K
  \tilde{\epsilon}\ ,
\end{equation}
where $g_{\alpha\beta}$ is the metric tensor, $R$ is the Ricci scalar
and $K$ is the trace of the extrinsic curvature. 
For a region bounded by two hypersurfaces $\Sigma_1$ and
$\Sigma_2$ the transition amplitude is formally given by
\begin{equation}\label{amp}
\langle g_2; \Sigma_2| g_1; \Sigma_1 \rangle =\int [Dg]
\exp{(i S[g])}\ .
\end{equation}
Here one integrates over all metric fields $g$ that smoothly interpolate between the
boundary fields $g_1$ and $g_2$.
If an intermediate 3-surface $\Sigma_3$ is introduced, one has
$S[g_{(23)}]+S[g_{(31)}] = S[g_{(21)}]$ where $g_{(ij)}$ interpolates
between $g_i$ and $g_j$ on $\Sigma_i$ and $\Sigma_j$, respectively.
The quantum-mechanical superposition principle implies
\begin{equation}
  \langle g_2; \Sigma_2| g_1; \Sigma_1 \rangle = \int [Dg_3]
\langle g_2; \Sigma_2| g_3; \Sigma_3 \rangle
\langle g_3; \Sigma_3| g_1; \Sigma_1 \rangle \ .
\end{equation}
The amplitude \eqref{amp} is only a formal expression and its
precise physical meaning is not
clear since the ``times'' $t_1$ and $t_2$ are merely coordinate
parameters. Despite much effort it has not been possible to decompose the metric
field into ``true dynamical degrees of freedom'' and some ``intrinsic time''; 
for a discussion and references, see
\cite{Wald:1984rg,Kiefer:2007ria}.

In the following we study the possibility to label the boundary
surfaces by values of a 3-form density
$A_{\alpha\beta\gamma}$, which is covariantly constant on a 3-surface.
Such a 3-form density can be sourced by the gravitational field, which
is achieved by equating its field strength to the canonical volume
density on $\mathcal{M}$. The corresponding action is obtained from
the Einstein-Hilbert action \eqref{hilbert} by adding a Lagrange
multiplier term,
\begin{equation}\label{hilbertA}
  S[g,A,\Lambda] = \int_{\mathcal{M}} (R\epsilon +
  \Lambda(dA-\epsilon))
  + 2\int_{\partial\mathcal{M}} K \tilde{\epsilon}\ ,
\end{equation}
where $\Lambda$ is an auxiliary scalar field. Note that the action is
invariant under the gauge transformation $A \rightarrow A + d\eta$
where $\eta$ is a 2-form field.
The equations of motion are obtained by varying the action with
respect to $g_{\alpha\beta}$, $A_{\alpha\beta\gamma}$ and $\Lambda$, which yields
\begin{align}
  G_{\alpha\beta} &= R_{\alpha\beta} - \frac{1}{2}g_{\alpha\beta}R =
                    - \frac{1}{2}\Lambda g_{\alpha\beta} \ , \label{einstein}\\
  \partial_\alpha \Lambda & = 0\ , \label{Lambda} \\
  4\partial_{[\alpha} A_{\beta\gamma\delta]}
                  &= \sqrt{g}
                    \epsilon_{\alpha\beta\gamma\delta} \ . \label{3vol}
\end{align}
Eqs.~\eqref{einstein} are Einstein's equations with a cosmological
term, Eq.~\eqref{Lambda}
implies that the scalar field $\Lambda$ becomes an unspecified
cosmological constant $\lambda$, and Eq.~\eqref{3vol} identifies the
field strength of $A$ with the canonical volume form.
The action \eqref{hilbertA} has been obtained by Henneaux and Teitelboim from
a constrained Hamiltonian analysis of a theory where the determinant
of the metric is treated as an external field \cite{Henneaux:1989zc}.
Instead of the 3-form density $A$ they used the dual
vector density, $A_{\alpha\beta\gamma}=\epsilon_{\delta\alpha\beta\gamma}\T^{\delta}$.

On a 3-surface $\Sigma_t$ the 3-form density $A$ is given by a
constant $A(t)$.  To study the time evolution
one has to
specify $g_{\alpha\beta}(t,x)$ and $A(t)$ on some initial 3-surface $\Sigma_1$,
together with a constant cosmological term, 
$\Lambda(t,x) = \lambda$. Einstein's equations then determine 
the metric at some later time $t_2$, and the integrated 3-form density at $t_2$ is given
by
\begin{equation}\label{volume}
  \A_2 = \A_1 + \mathcal{V}_{\mathcal{M}}[g] \ ,
\end{equation}
with
\begin{equation}\label{AVg}
  \A_t = A(t) \int_{\Sigma_t} d^3x \sqrt{h} \ , \quad
  \mathcal{V}_{\mathcal{M}}[g] = \int_{t_1}^{t_2}dt
  \int_{\Sigma_t}d^3x\sqrt{g} \ ,
\end{equation}
where $h$ is the induced volume density on $\Sigma_t$. By construction,
$\mathcal{A}_t$ increases monotonically with the coordinate time $t$.
This has motivated the interpretation of $\A_t$
as a ``cosmic time''
\cite{Henneaux:1989zc,Unruh:1988in,Unruh:1989db};
see, however, \cite{Kuchar:1991xd}.

Similar to Eq.~\eqref{amp} we can now consider transition amplitudes
where initial and final states depend on the fields $g$
and $A$. On the boundary surfaces $\Sigma_{1,2}$ the 3-form field $A$
is covariantly constant and can therefore be specified in terms of the integrals
$\mathcal{A}_{1,2}$. Hence, the transition amplitude takes the form 
\begin{align}\label{amplitude}
\langle g_2, &\A_2;\Sigma_2| g_1, \A_1; \Sigma_1 \rangle \nonumber\\
\quad &=\int [Dg][DA][D\Lambda]\exp{(i S[g,A,\Lambda])}
                       \nonumber\\
\quad  &  =\int [Dg][D\Lambda] \delta(\partial_\alpha\Lambda)
    \exp{\Big(i \Big(S[g] -\int_{\mathcal{M}} \Lambda\epsilon +
         \int_{\Sigma_2}d^3x \Lambda A - \int_{\Sigma_1} d^3x\Lambda
         A\Big)\Big)} \ ,
\end{align}
where $\delta(\partial_\alpha\Lambda) \equiv
\prod_{x,\alpha}\delta(\partial_\alpha\Lambda)$. 
Because of the $\delta$-function the integration over $\Lambda$ is
restricted to constant values. Assuming that this constant  is
fixed by boundary conditions we replace $\delta(\partial_\alpha\Lambda)$
by $\delta(\Lambda - \lambda_0)$, which leads to the
transition amplitude 
\begin{align}\label{ampA}
\langle g_2, \A_2;\Sigma_2| g_1, \A_1; \Sigma_1 \rangle 
   =\exp{\big(i \lambda_0 (\A_2 - \A_1)\big)}  
 \int [Dg] \exp{(i(S[g] - \lambda_0\mathcal{V}_{\mathcal{M}}[g])} \ .
\end{align}
Compared to standard GR the amplitude contains a phase
factor that is determined by the boundary conditions, and in the path
integral the Einstein-Hilbert action appears
with an undetermined cosmological constant $\lambda_0$,
which is the characteristic feature of unimodular gravity.
Contrary to the classical relation \eqref{volume} the integral
includes volumes that are not related to the boundary terms $\A_1$
and $\A_2$.
To obtain a better understanding of the boundary
conditions we now turn to the Hamiltonian formalism.
 
\section{The path integral in the ADM formalism}
\label{sec:ADM}
In the Arnowitt-Deser-Misner (ADM) \cite{Arnowitt:1962hi} formalism one starts 
from a foliation of the manifold $\mathcal{M}$  with spacelike
3-surfaces $\Sigma_t$. An embedding\footnote{We essentially follow the
  conventions of  the Lecture Notes on General Relativity by M. Blau
  (http://www.blau.itp.unibe.ch/GRLecturesnotes.html, 2021).}
of these 3-surfaces with coordinates $y^a$, $a=1,..,3$, into the ambient
space $\mathcal{M}$ is given by functions $x^\alpha(t,y^a)$, and the matrix 
$E^\alpha_a=\partial x^\alpha/\partial y^a \equiv \partial_a x^\alpha$ provides the
push-forward for tangent vectors of $\Sigma_t$ to tangent vectors of
$\mathcal{M}$. The metric induced on $\Sigma_t$ reads
\begin{equation}
  h_{ab} = g_{\alpha\beta}E^\alpha_a E^\beta_b \ ,
\end{equation}
and the vectorfield $E^\alpha_t \equiv t^\alpha$ represents the ``time
flow'' that can be decomposed into components
normal and tangential to $\Sigma_t$,
\begin{equation}
t^\alpha = \partial_t x^\alpha = N n^\alpha + E^\alpha_a N^a\ .
\end{equation}
Here $n^\alpha$ is a unit normal vector, $n^\alpha n_\alpha = -1$, and
$N$ and $N^a$ are the lapse function and the shift vector of the ADM
formalism, respectively. The induced metric $h^{\alpha\beta} =
E^{\alpha}_aE^{\beta}_b h^{ab}$,  lapse function and shift
vector determine the metric $g^{\alpha\beta}$ of the ambient space as
\begin{equation}\label{embedding}
  g^{\alpha\beta} = h^{\alpha\beta} - n^\alpha n^\beta
  =  E^{\alpha}_aE^{\beta}_b h^{ab} - \frac{1}{N^2}
  (t^\alpha - E^\alpha_a N^a)(t^\beta - E^\beta_b N^b) \ .
\end{equation}
The extrinsic curvature
\begin{equation}
  K_{ab} = E^{\alpha}_aE^{\beta}_b K_{\alpha\beta}\ , \quad
  K_{\alpha\beta} = h_\alpha^\gamma h_\beta^\delta \nabla_\gamma
  n_\delta\ ,
  \end{equation}
describes
the curvature of $\Sigma_t$ in the ambient space $\mathcal{M}$,
with $K = K^\alpha_\alpha = K^a_a = \nabla_\alpha n^\alpha$.

The Hamiltonian formalism for GR with a 3-from field
$A_{\alpha\beta\gamma}$, or equivalently the vector density
$\T^\alpha$, has previously studied in 
\cite{Kuchar:1991xd,Smolin:2009ti,Bufalo:2015wda}. In the following
discussion the emphasis lies on the effect of the boundary conditions.
In terms of the induced metric $h_{ab}$, the lapse function $N$, the
extrinsic curvature $K$, the field $\T^\alpha =(\T^t,\T^a)$ and $\Lambda$
the Lagrangian density $\mathcal{L}_g$ corresponding to
the action \eqref{hilbertA} reads,
\begin{equation}\label{Lg}
  \mathcal{L}_g = \sqrt{h} N (\tilde{R} + K_{ab} K^{ab} - K^2)
  + \Lambda (\partial_t \mathcal{T}^t  + \partial_a \mathcal{T}^a- \sqrt{h}N) \ . 
\end{equation}
Here $\tilde{R}$ is the Ricci scalar on $\Sigma_t$, which is
determined by $h_{ab}$ (see, for example, \cite{Wald:1984rg}).
The extrinsic curvature depends on
the time derivative of the metric $\dot{h}_{ab}=\partial_th_{ab}$,
\begin{equation}\label{extrinsic}
  K_{ab} = \frac{1}{2N}(\dot{h}_{ab} - D_{(a}N_{b)})\ .
\end{equation}
For the variables $h_{ab}$ and $\T^t$ one obtains the
canonical momenta
\begin{align}\label{momenta}
  \pi^{ab} = \sqrt{h}(K^{ab} - h^{ab} K)\ , \quad
   \pi_t = \Lambda \ .
\end{align}
The canonical momenta $\pi_a$, $\pi_\Lambda$, $\pi_N$ and $\pi_{N^a}$
for the variables $\T^a$, $\Lambda$, $N$ and $N^a$, respectively, 
all vanish. This leads to the Hamiltonian density 
\begin{align}\label{hgA}
   \mathcal{H}_g &= \pi^{ab}\dot{h}_{ab} +
  \pi_t\dot{\mathcal{T}}^t - \mathcal{L}_g \nonumber\\
  &= N(\mathcal{H} + \sqrt{h}\Lambda) + N^a\mathcal{H}_a
- \Lambda \partial_a \mathcal{T}^a\ ,
\end{align}
where 
\begin{equation}\label{Hdensities}
 \mathcal{H} = \sqrt{h}\left(-\tilde{R} + \frac{1}{h}\left(\pi^{ab}\pi_{ab} -
   \frac{1}{2}\pi^2\right)\right) \ , \quad
 \mathcal{H}_a = -2 \sqrt{h} D^b\left(\frac{1}{\sqrt{h}}\pi_{ab}\right)\ . 
\end{equation}
The fields $N$, $N^a$, $\T^a$ and $\Lambda$ are Lagrange multipliers.
Variation of the Hamiltonian $H_g = \int d^3x \mathcal{H}_g$ with
respect to these fields  yields the phase space constraints
\begin{equation}\label{constraints}
  \mathcal{H} + \sqrt{h}\Lambda = 0  \ , \quad \mathcal{H}_a = 0 \ , \quad
  \partial_a \Lambda = 0 \ , \quad \partial_a \mathcal{T}^a- \sqrt{h}N = 0 \ ,
\end{equation}
in agreement with the analysis in \cite{Smolin:2009ti}.

Using Eqs.~\eqref{momenta}, \eqref{hgA} and \eqref{constraints} we can
now write down the path integral. The third of the constraints
\eqref{constraints} implies that $\Lambda$ is spatially constant.
On the boundary 3-surfaces $\Sigma_{1,2}$ we can therefore specify 
constants $\lambda_{1,2}$.  On each 3-surface $\Sigma_t$ the field $\T^t$
can be split into a zero mode $A(t)$ and a field whose integral over
$\Sigma_t$ vanishes, $\T^t = A(t) + \partial_a\omega^a$. We can
therefore fix the gauge symmetry of the Lagrangian \eqref{Lg}, 
$\T^t \rightarrow \T^t - \partial_a\rho^a$, $\T^a \rightarrow \T^a + \partial_t\rho^a$,
by the condition $\partial_a\T^t = 0$.
On the boundary surfaces $\Sigma_{1,2}$ the 3-metric $h_{ab}$,
the constants $\A_t = \int d^3x \T^t = \int_{\Sigma_t} A$, and
$\lambda$ can be independently chosen, 
and the transition amplitude is given by the functional integral
\begin{align}\label{ampADM0}
\langle &h_2, \A_2, \lambda_2;\Sigma_2| h_1, \A_1, \lambda_1; \Sigma_1
          \rangle
          \nonumber\\
   &= \int [Dh_{ab}][D\pi^{ab}][D\T^t][D\pi_t][D\Lambda]
                [DN][DN^a][D\T^a]\delta(\pi_t - \Lambda)\delta(\partial_a\Lambda)
                 \delta(\partial_a\T^t)      \nonumber\\
  &\quad\quad\times\exp{\Big(i \int_{\mathcal{M}} d^4x(\pi^{ab}\dot{h}_{ab} +
  \pi_t\dot{\mathcal{T}}^t- 
         N(\mathcal{H} + \sqrt{h}\Lambda) - N^a\mathcal{H}_a
+ \Lambda \partial_a \mathcal{T}^a)\Big)} \ .
\end{align}
For spatially constant $\Lambda$ the exponent no longer depends
on $\T^a$,
and integration over the fields $\mathcal{T}^a$ yields a constant factor.
Performing the integration over $\pi_t$ and replacing
$\delta(\partial_a\Lambda)$ by $[D\lambda(t)]\delta(\Lambda - \lambda(t))$,
the amplitude becomes
\begin{align}\label{ampADM}
\langle h_2, &\A_2, \lambda_2;\Sigma_2| h_1,\A_1, \lambda_1; \Sigma_1
          \rangle
          \nonumber\\
  &=  \int [Dh_{ab}][D\pi^{ab}][D\A_t][D\lambda(t)][DN][DN^a]
                       \nonumber\\
             &\quad\quad\times\exp{\Big(i\int_{t_1}^{t_2}dt\lambda(t)\dot{\A_t}
               + i\int_{\mathcal{M}} d^4x(\pi^{ab}\dot{h}_{ab} -
               N(\mathcal{H}
               + \sqrt{h}\lambda(t)) - N^a\mathcal{H}_a)\Big)} \ .
\end{align}
After a partial integration yielding the boundary term
$[\lambda(t)\A_t]\big|_1^2$, the integral over $\A_t$ can be performed which leads
to a factor $\delta(\dot{\lambda}(t))$ in the functional integral. Since
$\lambda(t)$ has to satisfy the boundary conditions $\lambda(t_{1,2})
= \lambda_{1,2}$ we replace $\delta(\dot{\lambda}(t))$ by
$\delta(\lambda(t) - \lambda_1) \delta(\lambda(t) - \lambda_2)$.
Integrating over the canonical momenta $\pi^{ab}$ we finally obtain,
\begin{align}\label{amp2}
\langle h_2, \mathcal{A}_2,&\lambda_2;\Sigma_2| h_1, 
                                       \mathcal{A}_1, \lambda_1;
          \Sigma_1 \rangle  \nonumber\\
        & =\delta(\lambda_2-\lambda_1)
   \exp{(i\lambda_1(\A_2 - \A_1))}\mathcal{N}_2\mathcal{N}_1
               \int [Dg] \exp{(i S[g] -
          \lambda_1\mathcal{V}_{\mathcal{M}}[g])}  \ ,
\end{align}
where $\mathcal{N}_{1,2}$ are normalization factors related to the boundaries.
The amplitude essentially agrees with Eq.~\eqref{ampA}, with
the important difference that instead of an unspecified constant
$\lambda_0$ now the boundary values $\lambda_1$ and $\lambda_2$
appear. The result is consistent with the one obtained in \cite{Bufalo:2015wda}.
Note that the integral over the metric is not affected by the boundary
conditions $\A_{1,2}$. In particular the integration includes metric fields $g$
interpolating between $h_1$ and $h_2$ with volumes of arbitrary
size\footnote{The result differs from the path integral obtained in
  \cite{Smolin:2009ti} where the integration is restricted to volumes of
  some fixed size that is introduced via a gauge fixing condition.}.
The phase factor suggests that
$\A$ and $\lambda$ are conjugate variables with $\A$ and
$\lambda$ playing the role of ``time'' and ``energy'', respectively
\cite{Henneaux:1989zc}.
However, $\A$ can take arbitrary
positive and negative values and it does not increase monotonically
with the parameter time $t$. Therefore, generically, $\A$ cannot be
interpreted as a time parameter.
                                       
The amplitude clearly satisfies the superposition principle. Splitting the
manifold $\mathcal{M}_{(21)}$ bounded by $\Sigma_2$ and $\Sigma_1$
into two regions $\mathcal{M}_{(23)}$ and $\mathcal{M}_{(31)}$ separated
by $\Sigma_3$, one has
\begin{align}
  \int [Dh_3]&d\A_3 d\lambda_3
               \langle h_2, \mathcal{A}_2,\lambda_2;\Sigma_2| h_3, 
                               \mathcal{A}_3, \lambda_3; \Sigma_3 \rangle 
\langle h_3, \mathcal{A}_3,\lambda_3;\Sigma_3| h_1, 
               \mathcal{A}_1, \lambda_1; \Sigma_1 \rangle \nonumber \\
             &= \delta(\lambda_2-\lambda_1) \exp{(i\lambda_1(\A_2 - \A_1))}
               \mathcal{N}_2(\mathcal{N}_3)^2 \mathcal{N}_1
               \int [Dg_{(23)}][Dh_3][Dg_{(31)}] \nonumber\\
  &\hspace{2cm} \times\exp{(i(S[g_{(23)}]+S[g_{(31)}]
               -\lambda_1(\mathcal{V}_{\mathcal{M}}[g_{(23)}] +
               \mathcal{V}_{\mathcal{M}}[g_{(31)}]))} \nonumber\\
   &= \langle h_2, \mathcal{A}_2,\lambda_2;\Sigma_2| h_1, 
                               \mathcal{A}_1, \lambda_1; \Sigma_1
     \rangle \ , 
                 \end{align}
  where the metric $g_{(ij)}$ interpolates between $h_i$ and $h_j$ on $\Sigma_i$
  and $\Sigma_j$, respectively, and
the boundary normalization factors have been fixed to
$(\mathcal{N}_i)^{-2} = \int d\A_i$. 

In the semiclassical approximation the exponent in \eqref{amp2}  is evaluated at a
stationary point satisfying Einstein's equations,
\begin{equation}
 R_{\alpha\beta} - \frac{1}{2}g_{\alpha\beta}R =
                    - \frac{1}{2}\lambda_1 g_{\alpha\beta} \ .
\end{equation}
As a simple example consider the case of positve cosmological
constant, $\lambda_1 > 0$, for which a solution of Einstein's
equations is given by the FLRW metric $g_{\text{FLRW}}$ with an exponentially growing 
scale factor. For a foliation with 3-spheres one has
\begin{equation}\label{FLRW}
  ds^2 = - N(t)dt^2 + h_{ab}(t,y^a)dy^ady^b \ , \quad
  h_{ab}(t,y^a) = a(t)^2 \tilde{h}_{ab} \ ,
\end{equation}
where $a(t)$ is the scale factor and $\tilde{h}_{ab}$ is the well
known metric on the unit 3-sphere. Volume and Ricci scalar of the 3-sphere 
are given by
$\mathcal{V}_{\Sigma_t} = 2\pi^2 a(t)^3$ and $\tilde{R} = 6/a(t)^2$,
respectively. The four-dimensional Ricci scalar is $R=2\lambda_1$.
In Eq.~\eqref{FLRW} a comoving time coordinate has been
chosen, hence the shift vector $N^a$ is zero. The presence of the
lapse function allows for reparametrizations of time.

From Eqs.~\eqref{extrinsic} and \eqref{momenta}
one obtains for the extrinsic curvature and the canonical momenta
\begin{equation}
  K_{ab} = \frac{\dot{a}a}{N}\tilde{h}_{ab}\ , \quad
    \pi^{ab} = -2\frac{\dot{a}}{N}\tilde{h}^{ab} \ ,
\end{equation}  
and using Eqs.~\eqref{Hdensities} and \eqref{constraints}
with a cosmological constant $\lambda_1$ one finds
for the Hamiltonian constraint
\begin{equation}
  \mathcal{H} + \sqrt{h}\lambda_1 = 
  -6 \sqrt{\tilde{h}}a^3 \left(\left(\frac{\dot{a}}{Na}\right)^2
    + \frac{1}{a^2}- \frac{\lambda_1}{6}\right) = 0 \ ,
\end{equation}
which corresponds to Friedmann's equation. Einstein's equations also
yield Raychaudhuri's equation for the second time-derivative of the
scale factor, and
the two equations together have the well-known solution 
$a(\tau) = \sqrt{6/\lambda_1}\cosh{(\sqrt{\lambda_1/6} \tau)}$,
where $d\tau = N(t)dt$ determines the proper comoving time $\tau$.
Considering for simplicity
times $\tau\gg \sqrt{6/\lambda_1}$, one obtains for 
the total volume ($a_2\equiv a(\tau_2) \gg a(\tau_1) \equiv a_1$)
\begin{equation}\label{classicalvolume}
\mathcal{V}_{\mathcal{M}} = \int_{t_1}^{t_2} dt \int_{\Sigma_t} d^3x \sqrt{g}
= 2\pi^2 \int_{t_1}^{t_2} dt N(t) a(t)^3  \simeq 2\pi^2
\sqrt{\frac{2}{3\lambda_1}}a_2^3 \ .
\end{equation}
With $h_{ab}(t)$ determined by $a(t)$, 
the amplitude \eqref{amp2} can be written as
\begin{equation}\label{ampFRWA}
\langle a_2, \A_2,\lambda_2;\Sigma_2 | a_1, \A_1,
               \lambda_1; \Sigma_1 \rangle 
\propto \delta(\lambda_2-\lambda_1)
          \exp{\big( i \lambda_1(\A_2 - \A_1 + \mathcal{V}_{\mathcal{M}})\big)}\ .
        \end{equation}
Note that the action for the FLRW metric is given by
$S[g_{\text{FLRW}}] = \lambda_1\mathcal{V}_{\mathcal{M}}$.

\section{Unimodular gravity}
\label{sec:UR}
It is instructive to compare covariant UG with a 3-form density and
UG with the constraint $\sqrt{g} = \omega$, where $\omega$ is
some nondynamic background spacetime volume element. 
In this case one starts from the Hamiltonian density
\begin{align}\label{hgB}
  \mathcal{H}_g= N(\mathcal{H} + \sqrt{h}\Lambda) + N^a\mathcal{H}_a
- \Lambda \omega\ ,
\end{align}
where $\mathcal{H}$ and $\mathcal{H}_a$ are again given by Eq.~\eqref{Hdensities}
and $\Lambda$ is a Lagrange multiplier field. Variation with
respect to $N^a$ and $\Lambda$ yields the constraints
\begin{equation}\label{constraintsB}
  \mathcal{H}_a = 0 \ , \quad N\sqrt{h} - \omega = 0 \ .
\end{equation}
Because $N$ is now fixed to $\omega/\sqrt{h}$ there is no Hamiltonian
constraint. However, a tertiary constraint follows from
the requirement that the time evolution preserves the momentum
constraint. Using the Poisson bracket algebra
\begin{equation}
  \begin{split}
\{(h^{-1/2}\mathcal{H})(x),\mathcal{H}_a(x')\} &= \partial_a(h^{-1/2}\mathcal{H})(x))
\delta(x,x') \ ,\\
\{\mathcal{H}_a(x),\mathcal{H}_b(x')\} &= \mathcal{H}_a(x')
\partial_b\delta(x,x')  + \mathcal{H}_b(x') \partial_a\delta(x,x') \ ,
\end{split}
\end{equation}
one obtains the constraint
\begin{align}
 0 &= \left\{H_g,\int d^3x \xi^a\mathcal{H}_a\right\} =
 \left\{\int d^3x' (\omega h^{-1/2}\mathcal{H} + N^b\mathcal{H}_b),
                 \int d^3x \xi^a\mathcal{H}_a\right\} \nonumber \\
 &= \int d^3x \xi^a (\omega\partial_a(h^{-1/2}\mathcal{H}
- (\partial_aN^b+\partial_cN^c\delta_a^b)\mathcal{H}_b
- N^b\partial_b\mathcal{H}_a)  \ .\nonumber
\end{align}
For arbitary vector fields $N^a$ and $\xi^a$ this implies
\cite{Henneaux:1989zc,Unruh:1988in,Unruh:1989db}
\begin{equation}\label{constraintB}
\partial_a\left(\frac{1}{\sqrt{h}}\mathcal{H}\right) = 0 \ .
\end{equation}
The constraint can be solved by
\begin{equation}\label{hcB}
  \mathcal{H} + \sqrt{h}\lambda = 0 \ ,
\end{equation} 
where $\lambda$ is constant, which has to be satisfied on each
3-surface. Therefore we again have to specify constants
$\lambda_{1,2}$ on the boundary surfaces $\Sigma_{1,2}$.

It is now straightforward to write down the path integral for the
transition amplitude analogous to Eq.~\eqref{ampADM},
 \begin{align}
\langle h_2, \lambda_2;\Sigma_2| h_1, \lambda_1; \Sigma_1 \rangle 
  &=  \int [Dh_{ab}][D\pi^{ab}][DN^a][D\lambda(t)]
                \delta(\mathcal{H}+\sqrt{h}\lambda(t))\nonumber\\
   &\quad\quad\times\exp{\Big(i \int_{\mathcal{M}} d^4x(\pi^{ab}\dot{h}_{ab} 
  - h^{-1/2}\omega\mathcal{H} - N^a\mathcal{H}_a )\Big)} \ ,
 \end{align}
Note that the constraint \eqref{hcB} has been implemented for each
hypersurface $\Sigma_t$ and that the integration is performed over 
$\lambda(t)$, with the boundary conditions $\lambda(t_{1,2}) =
\lambda_{1,2}$.  Exponentiating the
constraint \eqref{hcB} by introducing again a Lagrange multiplier $N$,
and shifting $N$ to $N-\omega/\sqrt{h}$
one arrives at
\begin{align}\label{ampUGH}
\langle h_2, \lambda_2;&\Sigma_2| h_1, 
               \lambda_1; \Sigma_1 \rangle \nonumber\\
  \quad &= 
          \int [Dh_{ab}][D\pi^{ab}][DN][DN^a][D\lambda(t)] \nonumber\\
&\quad\quad\times\exp{\Big(i \int_{\mathcal{M}} d^4x(\pi^{ab}\dot{h}_{ab} 
  - N(\mathcal{H} + \sqrt{h}\lambda(t)) - N^a\mathcal{H}_a
+ \lambda(t)\omega)\Big)} \ .
\end{align}
We can now integrate over the canonical momenta $\pi^{ab}$ which yields
the amplitude in Lagrangian form,
\begin{align}\label{ampUGL}
\langle h_2, \lambda_2;&\Sigma_2| h_1, 
               \lambda_1; \Sigma_1 \rangle \nonumber\\
  \quad &= 
                \int [Dg][D\lambda(t)] \exp{\Big(iS[g] - i\int_{\mathcal{M}} d^4x
                 \lambda(t)(\sqrt{g} -\omega)\Big)}
          \ .  
\end{align}
Contrary to Eq.~\eqref{amp2} the amplitude does not contain a factor
$\delta(\lambda_1 - \lambda_2)$. Instead a Lagrange multiplier appears
for the volume of each 3-surface $\Sigma_t$. 
Correspondingly, integration over $\lambda(t)$ yields
 a product of $\delta$-functions in the functional integral,
 \begin{align}
\langle h_2, \lambda_2;\Sigma_2| h_1, 
               \lambda_1; \Sigma_1 \rangle 
   = \int [Dg] \prod_t\delta(N(t)\mathcal{V}_{\Sigma_t}-\Omega(t))
           \exp{(i S[g])} \ ,
\end{align}
 with
\begin{equation}\label{cV}
  \mathcal{V}_{\Sigma_t} = \int_{\Sigma_t} d^3x \sqrt{h} \ , \quad
         \Omega(t) = \int_{\Sigma_t} d^3x\omega \ \equiv \Omega(t) \ .
 \end{equation}
The spatially integrated background volume element $\Omega(t)$ depends 
on the chosen coordinate system.

The transition amplitude satisfies a Schr\"odinger equation with
respect to the upper end $t_2$ of the time integration. Using the
momentum constraint in Eq.~\eqref{constraintsB} and the 
constraint \eqref{hcB} one obtains for the Hamiltonian appearing in
the exponent of \eqref{ampUGH} at the boundary $\Sigma_2$,
\begin{equation}
 H_g\big|_{\Sigma_2} = \int_{\Sigma_2}  \big(N(\mathcal{H} +
 \sqrt{h}\lambda(t)) + N^a\mathcal{H}_a - \lambda(t)\omega\big)
= -\lambda_2\Omega(t_2) \ .
\end{equation}
This yields the Schr\"odinger equation
\begin{align}\label{SEUG1}
    i\frac{\partial}{\partial t_2}
 \langle h_2, &\lambda_2;\Sigma_2| h_1, \lambda_1; \Sigma_1 \rangle
= \langle h_2, \lambda_2;\Sigma_2| H_g\big|_{\Sigma_2} |h_1, \lambda_1; \Sigma_1
\rangle \nonumber\\
              & =-\lambda_2\Omega(t_2)
     \langle h_2, \lambda_2;\Sigma_2|h_1, \lambda_1; \Sigma_1\rangle \ .
\end{align}
Unruh and Wald obtained this equation for a wave function
 $\psi(t;h,\lambda)$, with $t$ playing the role of a ``Heraclitian time parameter''
 \cite{Unruh:1989db}. Note that $\Omega(t)$ can be
 absorbed into a redefined time variable.
 
In the semiclassical
approximation the amplitude \eqref{ampUGL} is dominated by the contribution of 
stationary points that satisfy the field equations
\begin{equation}
  R_{\alpha\beta} - \frac{1}{2} R g_{\alpha\beta}
  = -\frac{1}{2}\lambda(t)g_{\alpha\beta}\ , \quad \partial_\alpha \lambda(t) = 0 \ ,
\end{equation}
where the second equation follows from the Bianchi identity.
Variation with respect to $\lambda(t)$ yields
\begin{equation}\label{cV2}
  N(t) \mathcal{V}_{\Sigma_t} = \Omega(t) \ .
\end{equation}
Since for stationary points $\lambda(t)$ is constant, the amplitude is
again proportional to
$\delta(\lambda_2-\lambda_1)$. With $R = 2\lambda_1$, one finds
 \begin{equation}\label{ampUG}
\langle h_2, \lambda_2;\Sigma_2| h_1, \lambda_1; \Sigma_1 \rangle
\propto \delta(\lambda_2-\lambda_1)\exp{\big( 2 i\lambda_1 \int_{t_1}^{t_2} dt \Omega(t)\big)} \ .
\end{equation}

Consider now the example of the FLRW metric \eqref{FLRW}.
From Eqs.~\eqref{cV} and \eqref{cV2} one obtains the constraint
 \begin{equation}
   2\pi^2 N(t) a(t)^3 = \Omega(t) \ .
 \end{equation}
Knowing $a(t)$ from the solution of Friedmann's and Raychaudhuri's
equations, this fixes the lapse function, and therefore the time coordinate, to 
$N(t) = \Omega(t)/(2\pi^2a(t)^3)$. For an exponential expansion the
growth of $a(t)^3$ is compensated by the decrease of $N(t)$ such that
the amplitude is still given by Eq.~\eqref{ampUG}.

Finally, we compare the result of unimodular gravity with standard general
relativity. Here the transition amplitude reads
\begin{align}\label{ampGR}
\langle h_2;\Sigma_2| h_1; \Sigma_1\rangle  &= 
          \int [Dh_{ab}][D\pi^{ab}][DN][DN^a]
                       \nonumber\\
  &\quad\times\exp{\Big(i \Big(\int_{\mathcal{M}} d^4x(\pi^{ab}\dot{h}_{ab} - 
         N(\mathcal{H} + \sqrt{h}\lambda) - N^a\mathcal{H}_a\Big)\Big)} \ ,
\end{align}
where $\lambda$ is now a parameter of the Lagrangian. The
characteristic feature of the amplitude is the Hamiltonian and the
momentum constraints that follow from the integration over $N$ and $N^a$,
respectively,
\begin{equation}
  \mathcal{H} + \sqrt{h}\lambda = 0 \ , \quad \mathcal{H}_a = 0 \ .
\end{equation}
Hence, the amplitude satisfies the differential equation
\begin{align}\label{SEGR1}
    i\frac{\partial}{\partial t_2}
 \langle h_2;\Sigma_2| h_1; \Sigma_1 \rangle
&= \langle h_2;\Sigma_2| H_g\big|_{\Sigma_2} |h_1; \Sigma_1
\rangle \nonumber\\
             & =  \left(\int_{\Sigma_2} d^3x (N\mathcal{H} + N^a\mathcal{H}_a)\right)
 \langle h_2;\Sigma_2|h_1; \Sigma_1\rangle  = 0 \ .
\end{align}
This is the well-known feature of the Wheeler-DeWitt equation that
in general relativity wave functions have no time dependence.

In the semiclassical approximation one has to solve Einstein's
equation for a given cosmological constant $\lambda$.
The solution $g_{\text{cl}}$ yields
$R=2\lambda$ and an exponentially growing scale factor with 
spacetime volume $\mathcal{V}_{\mathcal{M}} = \simeq
\sqrt{2}2\pi^2 a_2^3/\sqrt{3\lambda}$.  The corresponding amplitude reads
\begin{equation}\label{WKB}
\langle h_2;\Sigma_2| h_1; \Sigma_1 \rangle
\propto \exp{\big(  i S[g_{\text{cl}}]\big)} =
\exp{\big(  i\lambda \mathcal{V}_{\mathcal{M}}\big)} \ .
\end{equation}

Unimodular gravity in the Henneaux-Teitelboim form shares features of
standard general relativity as well as
unimodular gravity with a fixed background volume element.  As
discussed above, the cosmological term is not a parameter of 
Lagrangian but appears as a boundary term, i.e., as a property of 
states. On the other hand, wavefunctions do not depend on time.
For the amplitude \eqref{ampADM} the constraints
\begin{equation}
\mathcal{H} + \sqrt{h}\lambda(t) = 0 \ , \quad \mathcal{H}_a = 0 
\end{equation}
hold on each 3-surface $\Sigma_t$. Hence, as in general relativity, the
Hamiltonian on the boundary surface $\Sigma_2$ vanishes,
$H_g\big|_{\Sigma_2} = 0$. This implies for the amplitude
\begin{equation}\label{SEHT1}
\begin{split}
  i\frac{\partial}{\partial t_2}
  \langle h_2, \A_2, &\lambda_2;\Sigma_2| h_1, \A_2, \lambda_1;
  \Sigma_1\rangle \\
  &= \langle h_2, \A_2, \lambda_2;\Sigma_2| H_g\big|_{\Sigma_2}
  |h_1, \A_2, \lambda_1; \Sigma_1 \rangle = 0 \ .
\end{split}
\end{equation}
This result is analogous to Eq.~\eqref{SEGR1}, with the only difference that
in addition to the metric also
the integrated 3-form field $\A$ and a cosmological constant $\lambda$
appear as variables of the boundary states.

\section{Time (in)dependent wave functions}
\label{sec:wavefunctions}
In quantum gravity there is no intrinsic time and therefore no
canonical formalism and no Hilbert space of physical states as in
quantum field theory in
flat spacetime. One considers wave functions of the form
\begin{equation}\label{WdW}
  \psi[h;\Sigma] = \int_{\mathcal{C}} [Dg] \exp{\big(iS[g]\big)}\ ,
\end{equation}
where $\mathcal{C}$ denotes a class of spacetimes with only one compact
spacelike 3-surface $\Sigma$ as boundary on which $h$ is the induced metric
\cite{Hartle:1983ai,Kiefer:2007ria}. The scalar product
\begin{equation}
  (\psi',\psi) = \int [Dh] \bar{\psi}'[h;\Sigma]\psi[h;\Sigma]
= \int _{(\mathcal{C'},\mathcal{C})} [Dg] \exp{\big(iS[g]\big)}
\end{equation}
has the geometric interpretation as a sum over all histories which lie
in class $\mathcal{C}$ to the past of the surface and in the time
reversed class $\mathcal{C'}$ to its future \cite{Hartle:1983ai}. This
product cannot be interpreted as a scalar product of physical states
in a Hilbert space. Only in the semiclassical approximation the 
WKB form \eqref{WKB} of the transition amplitude is reproduced. But
this is just classical physics and it is far from clear how to extend
the semiclassical approximation to the quantum regime. For the
  quantum mechanical system of a homogeneous scalar field in FLRW
  spacetime the scalar field can be used as a time variable \cite{Brunetti:2009eq}.

Since the interpretation of solutions of the Wheeler-DeWitt equation
is very challenging (see, for example,
\cite{Weinberg:1988cp,Unruh:1989db,Kiefer:2007ria,Hebecker:2021egx}), UG
appeared as an interesting possibility to achieve a canonical
quantization of gravity \cite{Unruh:1988in,Unruh:1989db}. Starting from orthogonal eigenstates
of the variables $h$ and $\lambda$ one can define time-dependent wave functions 
\begin{equation}
  \psi[h,\lambda;\Sigma_t] = \int [Dh_1] d\lambda_1
  \langle h, \lambda; \Sigma_t| h_1, 
               \lambda_1; \Sigma_1 \rangle \phi(h_1,\lambda_1)
\end{equation}
by integrating the transition amplitude over initial-state parameters
weighted with some distribution function $\phi$. Like the amplitude
\eqref{ampUGL} also the wave functions satisfy a
Schr\"odinger equation,
\begin{equation}\label{SEUG2}
  i\frac{\partial}{\partial t}\psi[h,\lambda;\Sigma_t]
  = - \lambda\Omega(t)\psi[h,\lambda;\Sigma_t] \ .
\end{equation}  
For these wave functions one can define a scalar product by integrating
over the variables $h$ and $\lambda$,
\begin{equation}
  (\psi', \psi) =
  \int [Dh] d\lambda \bar{\phi'}(h,\lambda)\phi(h,\lambda) \ .
\end{equation}
Hence, normalizable states can be defined
such that a probability interpretation of
$|\psi(t)|^2 \equiv (\psi[h,\lambda;\Sigma_t],\psi[h,\lambda;\Sigma_t])$ is
possible, which is difficult to achieve for solutions of the
Wheeler-DeWitt equation. The transition amplitude of the theory is given by
Eq.~\eqref{ampUGL}. As discussed in the previous section it has the
characteristic feature that the cosmological constant enters as a property
of states. On the other hand, the dependence of the time evolution of
states on an arbitrary background volume element appears as a weakness
of this modification of GR \cite{Anderson:1971pn}.

The Henneaux-Teitelboim version of UG is generally
covariant. Hence, as discussed above, wave functions are time
independent, as in GR. However, as in UG, the 3-form
field $A$, sourced by the metric $g$, leads to the appearance of a
cosmological constant as boundary term. From Eqs.~\eqref{amp2}
and \eqref{WdW} we infer that the wave function has the form
\begin{align}
  \psi[h,\A,\lambda;\Sigma] &= \int_{\mathcal{C}}
  [D\mu(\Sigma')][Dh']d\A'd\lambda'[Dg]
\langle h, \A,\lambda;\Sigma| h', \A', \lambda';\Sigma'\rangle
\phi(h',\A',\lambda';\Sigma') \nonumber\\
&= \int_{\mathcal{C}}[D\mu(\Sigma')]
    [Dh']d\A' d\lambda'[Dg]\delta(\lambda-\lambda')\exp{\big(iS[g]\big)} \nonumber\\
 &\hspace{1cm}  \times\exp{\big(i\lambda'(\A - \A' - \mathcal{V}_{\mathcal{M}}[g] )\big)}
                                 \phi(h',\A',\lambda';\Sigma')\  , 
\end{align}
where $\mathcal{C}$ again denotes a class of spacetimes with 
final 3-surface $\Sigma$ and initial 3-surfaces $\Sigma'$ over
which one integrates with some measure, 
$\mathcal{V}_{\mathcal{M}}[g]$ is the volume bounded by $\Sigma$ and
$\Sigma'$, and $\phi$ defines the initial states.
The wave function satisfies
a Schr\"odinger-type differential equation,
\begin{equation}
  i\frac{\partial}{\partial\A} \psi[h,\A,\lambda;\Sigma] =
  -\lambda \psi[h,\A,\lambda;\Sigma] \ ,
\end{equation}  
which is a consequence of the particular form of the action
\eqref{hilbertA}.\footnote{We could have started from an action where the Lagrange
multiplier term $\Lambda dA$ in Eq.~\eqref{hilbertA} is replaced by
$- Ad\Lambda$ \cite{Fiol:2008vk}, without changing the classical
equations of motion.  In this case the phase factor in
Eqs.~\eqref{ampA} and \eqref{amp2} disappears, the 3-form field can
be completely integrated out, the amplitudes in UR and GR are
identical, and the cosmological term is simply a constant
determined by initial conditions. This has been pointed out in
\cite{Padilla:2014yea}. However, in this version of the
theory the relation \eqref{volume} of the classical theory cannot
be obtained in a semiclassical approximation of the quantum theory.}
Note, however, that $\A$ is just the value of the 3-form field
 $A$ on the 3-surface $\Sigma$, with positive or negative
values, which generically cannot be interpreted as a time variable.
Only in a stationary-phase approximation the situation changes.
Then $S[g]-\lambda\mathcal{V}_{\mathcal{M}}[g]$ is evaluated for
solutions $g_{\text{cl}}$ of Einstein's equations with cosmological constant $\lambda$.
Moreover, stationarity of the phase with respect to $\lambda'$ yields the
relation \eqref{volume} between the boundary terms $\A$, $\A'$ and the volume
$\mathcal{V}_{\mathcal{M}}[g_{\text{cl}}]$,
\begin{equation}
  \A' = \A - \mathcal{V}_{\mathcal{M}}[g_{\text{cl}}] \ .
\end{equation}
Hence, in this approximation $\A$ increases monotonically with the parameter
time labeling the 3-surfaces of the foliation and can therefore be
used as a time variable.
A solution of Einstein's equation determines $h'$ as function of $h$ and
$\lambda$, and $\A'$ as function of $\A$, $h$ and $\lambda$.
Therefore, in the stationary-phase approximation the wave function becomes
\begin{align}
  \psi[h,\A,\lambda;\Sigma] 
\sim \int_{\mathcal{C}}
    [D\mu(\Sigma')] \exp{\big(iS[g_{\text{cl}}]\big)}
    \phi(h'[h,\lambda],\A - \mathcal{V}_{\mathcal{M}}[h,\lambda],\lambda;\Sigma') \ .
\end{align}
This means that the wave function at ''time'' $\A$ is obtained by integrating over
initial values at ``times'' $\A' < \A$.

\section{Summary and conclusions}
\label{sec:conclusion}

In the previous sections we have compared the path integral for
transition amplitudes in general relativity with the corresponding
amplitudes in the two versions of
unimodular gravity, the one with a nondynamical background volume element
and the covariant form with a 3-form gauge field. The
amplitude \eqref{amp2} for covariant UG agrees with the one of GR
except for a phase factor that depends on the boundary states and the
interpretation of the cosmological constant which is a property of the
boundary states rather than a parameter of the action. On the contrary, 
the amplitude \eqref{ampUGL} for UG with a background volume form
explicitly depends on the volume form $\omega$. Hence, the
two versions of UG generically lead to different predictions for
observables.

As covariant theories wave functions in GR and in covariant UG have no
time dependence and satisfy a Wheeler-DeWitt equation, which makes
their interpretation challenging, except for cases where a
semiclassical approximation applies. On the other hand,  UG with a
background volume form has a time variable that is canonically conjugate to 
the cosmological constant. Wave functions do depend on time and
satisfy a Schr\"odinger equation. It is interesting that in covariant
UG the 3-form gauge field integrated over spacelike hypersurfaces 
emerges as a ``cosmic time'' in the semiclassical approximation.

The change of the cosmological constant from a parameter of the action
to a property of states does not solve the cosmological constant
problem, but it does change it in a suggestive way
\cite{Weinberg:1988cp}, from a question of fine-tuning to a question
of initial conditions.
In general, a cosmological initial state is now a superposition of
states with different cosmological constants. 
It has been suggested that a vanishing or very small cosmological constant today
can be explained in such a framework, based on Euclidean quantum
gravity \cite{Baum:1983iwr,Hawking:1984hk,Coleman:1988tj} or,
alternatively, on unimodular gravity \cite{Ng:1990rw,Smolin:2009ti}.
It is interesting that the additional fields needed in unimodular
gravity occur in higher-dimensional supergravity theories
and in string theory \cite{Aurilia:1980xj,Hebecker:2021egx,Kaloper:2022utc}.


\subsection*{Acknowledgments}
We thank Klaus Fredenhagen for a helpful discussion and Marc Henneaux
for comments on the manuscript.

\appendix

\section{BRST quantization}
\label{sec:brst}

In this appendix we briefly review gauge fixing for the gravitational field,
which we have ignored in the previous sections.  
In covariant theories, such as general relativity or the
Hennaux-Teitelboim version of unimodular gravity, this is well known.
One may choose, for instance,  the harmonic gauge condition,
or de Donder gauge,
\begin{equation}
  C_\mu(g) = -\frac{1}{\sqrt{g}}g_{\mu\nu}
  \partial_\lambda(\sqrt{g} g^{\nu\lambda}) = 0 \ ,
\end{equation}
together with eight real Faddeev-Popov vector ghosts $u^\mu$ and
$\bar{u}^\mu$ for which the BRST invariance and the
unitarity of the physical S-matrix have been explicitly demonstrated
\cite{Kugo:1978rj,Kugo:1979gm}.

In unimodular gravity with a fixed background spacetime volume element
(we choose $\sqrt{g} = 1$) one can choose 
\begin{equation}\label{gf1}
  C(g) = \sqrt{g} - 1 = 0 
 \end{equation} 
as one of four gauge fixing conditions. A complete gauge fixing is
achieved by demanding in addition that the vector field $C_\mu$ is the gradient
of an auxiliary scalar field \cite{Buchmuller:1988yn},
\begin{equation}\label{gf2}
  C_\mu(g) + \partial_\mu B = 0 .
 \end{equation}.
 From Eqs.~\eqref{gf1} and \eqref{gf2} one obtains the gauge fixing
 Lagrangian\footnote{Compared to \cite{Buchmuller:1988yn} we have rescaled
  $\beta \rightarrow \alpha\beta$; moreover, since $\delta \neq 0$ only leads to
  an uninteresting variation of the harmonic gauge, we have set
  $\delta = 0$ for simplicity.}
\begin{equation}\label{gfL1}
  \mathcal{L}_{\text{GF}} = \frac{1}{2\alpha} \Lambda^\mu\Lambda_\mu
  - \Lambda^\mu\left(C_\mu + \beta \pd_\mu B\right)
  +\frac{1}{2\gamma}\Lambda^2 - \Lambda C \ ,
\end{equation}
where $\Lambda_\mu$ and $\Lambda$ are additional auxiliary fields.
The BRST invariant extension of the gauge fixing Lagrangian requires
two scalar ghosts $v$ and $\bar{v}$ in addition to the eight vector
ghosts $u^\mu$ and $\bar{u}^\mu$. The ghost lagrangian reads
\begin{equation}
  L_{\text{GH}} = -i(\bar{u}^\mu
  sC_\mu+\beta\bar{u}^\mu\partial_\mu v + \bar{v}sC ) \ , 
\end{equation}
where $s$ is a real, nilpotent antiderivation, and the BRST
transformations of all fields are given by
\begin{equation}
  \begin{split}
sg_{\mu\nu} &= u^\lambda\partial_\lambda g_{\mu\nu} + \partial_\mu
u^\lambda g_{\lambda\nu} +  \partial_\nu u^\lambda g_{\mu\lambda} \ ,\\
su^\mu &= u^\lambda\partial_\lambda u^\mu\ , \\
sB &= v\ , \quad sv = 0 \ ,\\
s\bar{u}^\mu &= i\Lambda^\mu\ , \quad s\Lambda^\mu = 0 \ ,\\
s\bar{v} &= i \Lambda\ , \quad s\Lambda = 0 \ .
\end{split}
\end{equation}
Recently, the fields $B$, $v$, $\bar{v}$ and $\Lambda$ have been
identified as a BRST quartet \cite{Baulieu:2020obv} and the
decoupling of BRST quartets in momentum space has been discussed in detail in
\cite{Kugo:2021bej}.

Eliminating in Eq.~\eqref{gfL1} the Lagrange
multiplier fields by their equations of
motion one obtains the gauge fixing Lagrangian 
\begin{equation}\label{gfL2}
 \mathcal{L}_{\text{GF}} = -\frac{\alpha}{2}\left(C^\mu + \beta
   \pd^\mu B\right)\left(C_\mu + \beta \pd_\mu B\right)
  - \frac{\gamma}{2} C^2 \ .
\end{equation}
In the linear approximation around flat space, $g_{\mu\nu} =
\eta_{\mu\nu} + h_{\mu\nu}$, the Green's functions can be written in a
compact form \cite{Buchmuller:1988yn}. With $\omega = (u_\mu,v)$
one finds for the ghost propagator matrix
\begin{equation}\label{propg}
  \langle \omega(x)\bar{\omega}(y) \rangle = -
 \begin{pmatrix}
                  \eta_{\mu\nu} - \frac{\pd_\mu \pd_\nu}{\Box} &  \pd_\mu\\
                  \frac{1}{\beta}\pd_\mu  & -\frac{1}{\beta}\Box
                   \end{pmatrix}
                                             \frac{1}{\Box} \delta^4(x-y) \ .
\end{equation}      
Correspondingly, defining for the gravitational field and the auxiliary scalar
field $\hat{h}=(h_{\mu\nu},B)$, one obtains the propagator matrix
\begin{equation}\label{propG}
 \begin{split}
  \langle \hat{h}(x)\hat{h}(y) \rangle &=
  \begin{pmatrix}
    D^{(\alpha)}_{\mu\nu\lambda\tau} & -\frac{1}{2\beta}(\eta_{\mu\nu}
    - \frac{4}{\gamma}(\Box + \gamma)\frac{\pd_\mu\pd_\nu}{\Box})\\
     -\frac{1}{2\beta}(\eta_{\lambda\tau} - \frac{4}{\gamma}(\Box +\gamma)
     \frac{\pd_\lambda\pd_\tau}{\Box})  &
     -\frac{1}{\beta^2\gamma}(\Box + \gamma(\frac{3}{2} - \frac{1}{\alpha}))
   \end{pmatrix}
  \\
  & \hspace{1cm} \times\frac{i}{\Box} \delta^4(x-y) \ , \\
\text{with} \hspace{.85cm}    D^{(\alpha)}_{\mu\nu\lambda\tau} &=
P^{(\alpha)}_{\mu\nu\lambda\tau} +
\frac{1}{\Box}(\eta_{\mu\nu}\pd_\lambda\pd_\tau +\eta_{\lambda\tau}\pd_\mu \pd_\nu)
 \\
    & \hspace{.5cm}
    - \frac{4}{\gamma} \left(\Box + \gamma\left(\frac{1}{2}
        +\frac{1}{\alpha}\right)\right)
    \frac{1}{\Box^2}\pd_\mu\pd_\nu\pd_\lambda\pd_\tau \ ,
   \\
\text{and} \hspace{1cm}  P^{(\alpha)}_{\mu\nu\lambda\tau} &=  \frac{1}{2}
  (\eta_{\mu\lambda}\eta_{\nu\tau} +
    \eta_{\mu\tau}\eta_{\nu\lambda} - \eta_{\mu\nu}\eta_{\lambda\tau})
  \\
  & \hspace{.5cm} -\frac{1}{2}\left(1-\frac{2}{\alpha}\right)\frac{1}{\Box}(
  \pd_\mu\pd_\lambda\eta_{\nu\tau} + \pd_\nu\pd_\lambda\eta_{\mu\tau}
  + \pd_\mu\pd_\tau\eta_{\nu\lambda} +
  \pd_\nu\pd_\tau\eta_{\mu\lambda}) \ .
\end{split}
\end{equation}
Note that $\Delta^{(\alpha)}_{\mu\nu\lambda\tau}(x-y) =
P^{(\alpha)}_{\mu\nu\lambda\tau}\frac{i}{\Box}\delta^4(x-y)$ is the well-known
graviton propagator in harmonic gauge.

The propagators in Eqs.~\eqref{propg} and \eqref{propG} involve terms
with $1/\Box^2$ and $1/\Box^3$.
The situation is similar for the propator matrix obtained
from the Lagrangian \eqref{gfL1} for the fields $h_{\mu\nu}$,
$B$, $\Lambda_\mu$ and $\Lambda$ \cite{Kugo:2021bej}. 
It is a non-trivial task to count the physical states for such a
system of propagators. In principle one has to rewrite the Lagrangian
in terms of simple-pole fields. An analysis directly in terms of
multiple-pole fields leads to the conclusion that the propagator
matrix decribes indeed just two physical graviton states with
helicities $\pm 2$ \cite{Kugo:2021bej}. As an alternative, the BRST
quantization in unimodular gauge has also been discussed using
ghost systems that include antisymmetric tensor fields
\cite{Kugo:2022iob,Kugo:2022dui}.


\bibliographystyle{JHEP}
\bibliography{unimodular}

\end{document}